\documentclass[preprint,number,sort&compress]{elsarticle}
\usepackage{graphicx}
\usepackage{epsfig}
\usepackage{amsmath}
\usepackage{amssymb}
\usepackage{lineno}
\usepackage{subfigure}

\begin{document}
\linenumbers

\title{Background Rejection in the DMTPC Dark Matter Search Using Charge Signals}
\author[mit,lns]{J.P.~Lopez\corref{cor1}}
\ead{jplopez@mit.edu}
\author[mit,lns]{D.~Dujmic}
\author[bu]{S.~Ahlen}
\author[mit,bmc]{J.B.R.~Battat\fnref{fn1}}
\author[mit,lns]{C.~Deaconu}
\author[mit,lns,mki,isn]{P.~Fisher}
\author[mit,lns]{S.~Henderson}
\author[bu]{A.~Inglis}
\author[mit,lns]{A.~Kaboth}
\author[mit,rhul]{J.~Monroe}
\author[mit,brandeis]{G.~Sciolla\fnref{fn1}}
\author[bu]{H.~Tomita}
\author[brandeis]{H.~Wellenstein}
\author[mit]{R.~Yamamoto}
\cortext[cor1]{Corresponding Author}
\fntext[fn1]{Present Address}

\address[mit]{Physics Department, Massachusetts Institute of Technology; Cambridge, MA 02139, USA}
\address[lns]{Laboratory for Nuclear Science, Massachusetts Institute of Technology; Cambridge, MA 02139, USA}
\address[bu]{Physics Department, Boston University; Boston, MA 02215, USA}
\address[bmc]{Physics Department, Bryn Mawr College; Bryn Mawr, PA 19010, USA}
\address[mki]{MIT Kavli Institute for Astrophysics and Space Research, Massachusetts Institute of Technology; Cambridge, MA 02139, USA}
\address[isn]{Institute for Soldier Nanotechnology, Massachusetts Institute of Technology; Cambridge, MA 02139, USA}
\address[brandeis]{Physics Department, Brandeis University; Waltham, MA 02453, USA}
\address[rhul]{Physics Department, Royal Holloway, University of London;  Egham, TW20 0EX, UK}

\begin{abstract}
The Dark Matter Time Projection Chamber (DMTPC) collaboration is developing a low pressure gas TPC for detecting Weakly Interacting Massive Particle (WIMP)-nucleon interactions. 
Optical readout with CCD cameras allows for the detection of the daily modulation of the direction of the dark matter wind.  
In order to reach sensitivities required for WIMP detection, the detector needs to minimize backgrounds from electron recoils.
This paper demonstrates that a simplified CCD analysis achieves $7.3\times10^{-5}$ rejection of electron recoils while a charge analysis yields an electron rejection factor of $3.3\times10^{-4}$ for events with $^{241}$Am-equivalent ionization energy loss between 40~keV and 200~keV. A combined charge and CCD analysis yields a background-limited upper limit of $1.1\times10^{-5}$ (90\% confidence level) for the rejection of $\gamma$ and electron events.
Backgrounds from alpha decays from the field cage are eliminated by introducing a veto electrode that surrounds the sensitive region in the TPC.
CCD-specific backgrounds are reduced more than two orders of magnitude when requiring a coincidence with the charge readout.
\end{abstract}

\maketitle

\section{Introduction}

In recent years, dark matter direct detection experiments have obtained seemingly contradictory results, both supporting the existence of WIMP dark matter~\cite{CogentPRL, CresstEPJ,DamaEPJ} and setting ever more stringent limits on its interaction cross section with nucleons~\cite{Xenon100PRL,CdmsPRL}. The tension between these experimental results highlights the need for a detection strategy that provides an unambiguous measurement capable of distinguishing between WIMP-nucleus scattering events and background nuclear recoils. The strong expected directional signature of WIMP-induced recoils, due to the motion of the earth through the galactic dark matter halo, may provide such an unambiguous evidence of WIMP-nucleus scattering~\cite{Spergel,LewinSmith}. A number of experimental groups are developing detectors to search for this signal~\cite{Ahlen:2009ev}, and techniques have been developed to analyze data with directional information to extract the possible directional signal of WIMP dark matter~\cite{Green:2006cb,Morgan:2005sq,Morgan:2004ys,PhysRevD.82.055011,PhysRevD.85.035006,PhysRevD.83.075002}.

The DMTPC collaboration uses a low-pressure time projection chamber (TPC) to search for WIMP-nucleon elastic scattering from WIMPs in the local dark matter halo.  
The TPC is filled with CF$_4$ gas to take advantage of the expected favorable WIMP-$^{19}$F spin-dependent cross section~\cite{19FScatter}.  
The WIMP interaction signature is a low-momentum nuclear recoil that leaves an ionization trail in the detector.
Primary-ionization electrons from nuclear recoils are amplified~\cite{Dujmic:2008ut} and the scintillation light from these avalanches is imaged by a charged-coupled device (CCD) camera.
Using the shape of the ionization trail, DMTPC detectors are able to identify the direction of the nuclear recoil. The analysis of CCD tracks is described in~\cite{10LPaper}. 

Electrons from $\beta$ decays and processes such as Compton scattering from $\gamma$- and x-rays are typically important backgrounds in dark matter direct detection sources. 
These can be due to radioactive contaminants in the materials used to construct the detector, radioactive components of the target material, and radioactive material in the laboratory environment. 
Such events are often rejected by fiducialization of the active volume. In experiments measuring the energy loss due to multiple processes, such as ionization, scintillation, and phonon excitation, these can further be identified by the relative fraction of energy loss due to each process~\cite{Xenon100PRL,CdmsPRL}. Other experiments attempt to create detectors that are wholly insensitive to electron recoils~\cite{Behnke15022008,Archambault2009185}.
Many of these techniques are inapplicable to current directional detector designs, which typically measure only the ionization component of recoil energy loss. Directional detection experiments can instead use the low density of ionization from electron recoils to identify and reject these events \cite{Burgos2007409,Mimac}. 

This work describes the charge readout systems of a prototype DMTPC detector and the corresponding analysis used to evaluate recoil properties such as position, energy, and geometry.  
It presents several studies taken with this detector in a surface laboratory at MIT.  
Section \ref{sec:readout} describes the charge readout systems of this detector and Section \ref{sec:reconstruction} describes event reconstruction.  

In DMTPC detectors, the CCD readout is used to reconstruct the ionization density projected onto a two-dimensional readout plane. The stopping power of electrons typically falls within the CCD read noise and is too low for these events to be reconstructed, while the stopping power of $\alpha$ particles and nuclear recoils is sufficient to accurately measure recoil properties.
The charge signals, which are sensitive to the total energy loss due to ionization, are able to measure and identify electronic recoils.  Section \ref{sec:emrej} describes a study using charge signals to determine the ability of a DMTPC detector to reject electronic recoils and to enhance the ability of the detector to reject these events through pulse shape analysis.

For DMTPC detectors, a number of backgrounds from CCD readout artifacts and interactions in the CCD bulk must also be eliminated in order to perform a low-background WIMP search.  Section \ref{sec:ccdrej} describes these backgrounds and rejection strategies using the charge signals. It also discusses the use of a veto region to identify and reject alpha decays from the outside the sensitive region. In certain circumstances, $\alpha$ particles can be misinterpreted by the CCD as nuclear recoils.

\section{Detector Design}
\begin{figure}
\centering
\includegraphics[width=\textwidth]{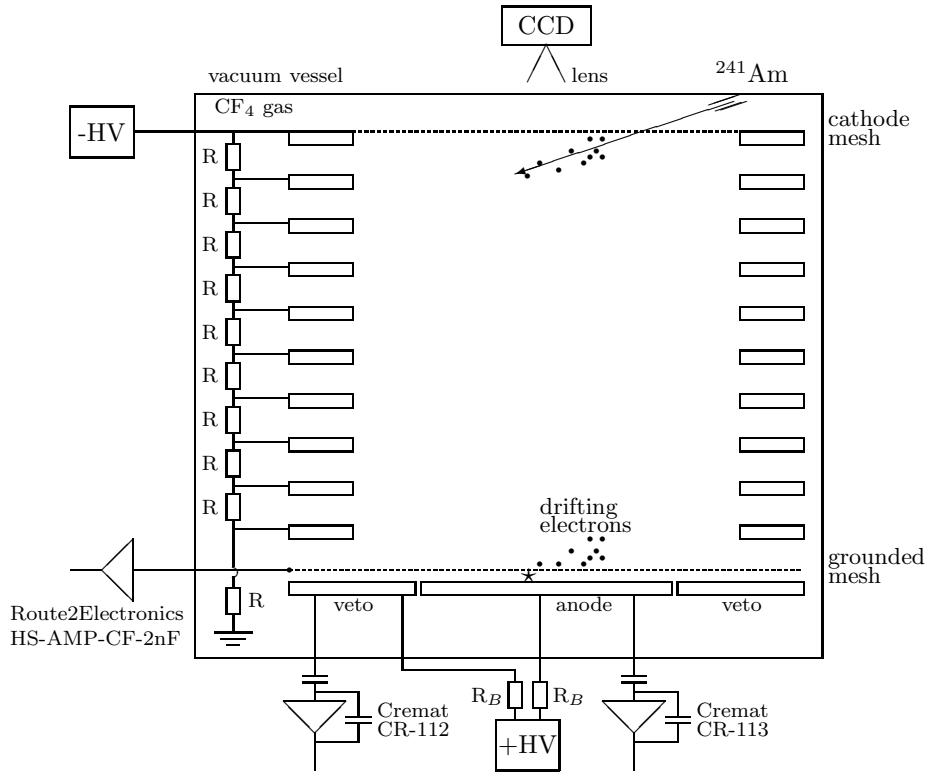}
\caption{A schematic of the detector: the drift field is created by a cathode mesh, field-shaping rings attached to a resistor chain and a ground mesh. The drift cage has a height of 10~cm and a diameter of 27~cm.
Primary ionization from a recoiling nucleus is drifted down to the ground mesh. The high-field amplification region is formed by the ground
mesh and the anode plane.  The grounded mesh is read  out with a fast amplifier and the veto and anode are read out with charge-sensitive preamplifiers. 
The central anode region has a diameter of 24.7~cm.
Scintillation light from the amplification region is recorded with the CCD camera.
\label{Detector_Drawing}}
\end{figure}

The data in this paper were taken with a detector in a surface lab in Cambridge, Massachusetts using a 75~Torr CF$_4$ gas target.  Figure \ref{Detector_Drawing} shows a schematic of this detector.  The field cage of the detector has a drift length of 10~cm with a wire mesh cathode held at -1.2~kV.  The drift cage is constructed of copper rings with a 27~cm inner diameter.  The amplification region consists of a stainless steel woven mesh separated by nonconductive spacers from an anode plane made of copper-clad G10.  The amplification mesh is coupled to an amplifier with a 30 $\Omega$ impedance to ground while the anode is biased at 680~V.  The separation is 440~$\mu$m.  

The scintillation light from electron amplification goes through a viewport at the top of the vacuum vessel and into a Nikon f/1.2 lens with a focal length of 55 mm. The lens directs the light into an Apogee Alta U6 CCD camera, which uses a Kodak KAF-1001 1024 x 1024 pixel CCD.  The 24~$\mu$m x 24~$\mu$m pixels are binned on-chip into 4 x 4 blocks prior to digitization, resulting in a 256 bin x 256 bin image.  The CCD images a 16.7~cm x 16.7~cm square centered on the circular anode.  The remainder of the anode is not imaged.
Each 4 x 4 pixel bin reads out a 650~$\mu$m x 650~$\mu$m region of the anode.  
An americium-241 $\alpha$ source was was used to determine a CCD energy calibration of 13.1$\pm$0.1 analog-to-digital units (ADU) per keV of energy loss contributing to ionization. The energy of the $\alpha$ particles from the source is attenuated by a thin film, and the mean energy was measured to be 4.44~MeV using an Ortec ULTRA ion-implanted-silicon detector \cite{HaykThesis}. Calculations from SRIM \cite{SRIM} show that over 99\% of the total energy loss of $\alpha$ particles at these energies contributes to ionization at this energy.
Measured recoil energies in this paper are reported in $\alpha$-equivalent energy units, denoted keV$_\alpha$.  Because the energy loss is dominated by ionization, the $\alpha$-equivalent energy will be similar to the more standard electron-equivalent energy (in units of keV$_{ee}$), which is difficult to determine in a detector that is largely insensitive to electrons.

\subsection{Charge Readout Systems}
\label{sec:readout}

The anode plane has a diameter of 26.7~cm and is separated into two regions. The outermost 1~cm, called the veto region, is used to identify ionization events occurring near the field cage rings. The inner 24.7~cm diameter region, called the anode region, is used to measure the energies of recoil events.
Charge signals from the anode region are amplified by a Cremat CR-113 charge sensitive preamplifier (CSP) \cite{Cremat}, while the signals from the veto region are amplified by a Cremat CR-112 CSP.  
The gain of the CR-113 CSP is 1.5~mV/pC, and that of the CR-112 CSP is 13 mV/pC.  Both have a nominal rise time of 20 ns when disconnected from the detector and a decay time of 50 $\mu$s.  
Recoil events have a typical rise time of approximately 1 $\mu$s due to the ion drift velocity across the amplification region, so the peak output voltage of the anode CSP gives a very accurate measurement of the total track ionization.  
 Low-energy $\alpha$ particles from the $^{241}$Am run described in Section \ref{sec:select}, with 40~keV$_\alpha<E<400$~keV$_{\alpha}$, give an energy calibration of 0.251$\pm$0.003~mV per keV$_\alpha$ in the central anode channel.  There is a linear relationship between the energies measured in charge and light in this energy range.

\begin{figure}
\centering

   \subfigure[Mesh (left), central anode (right, dark), and veto (right, light) signals from a 75~keV$_\alpha$ $\alpha$ track]{
   \includegraphics[width=0.45\textwidth]{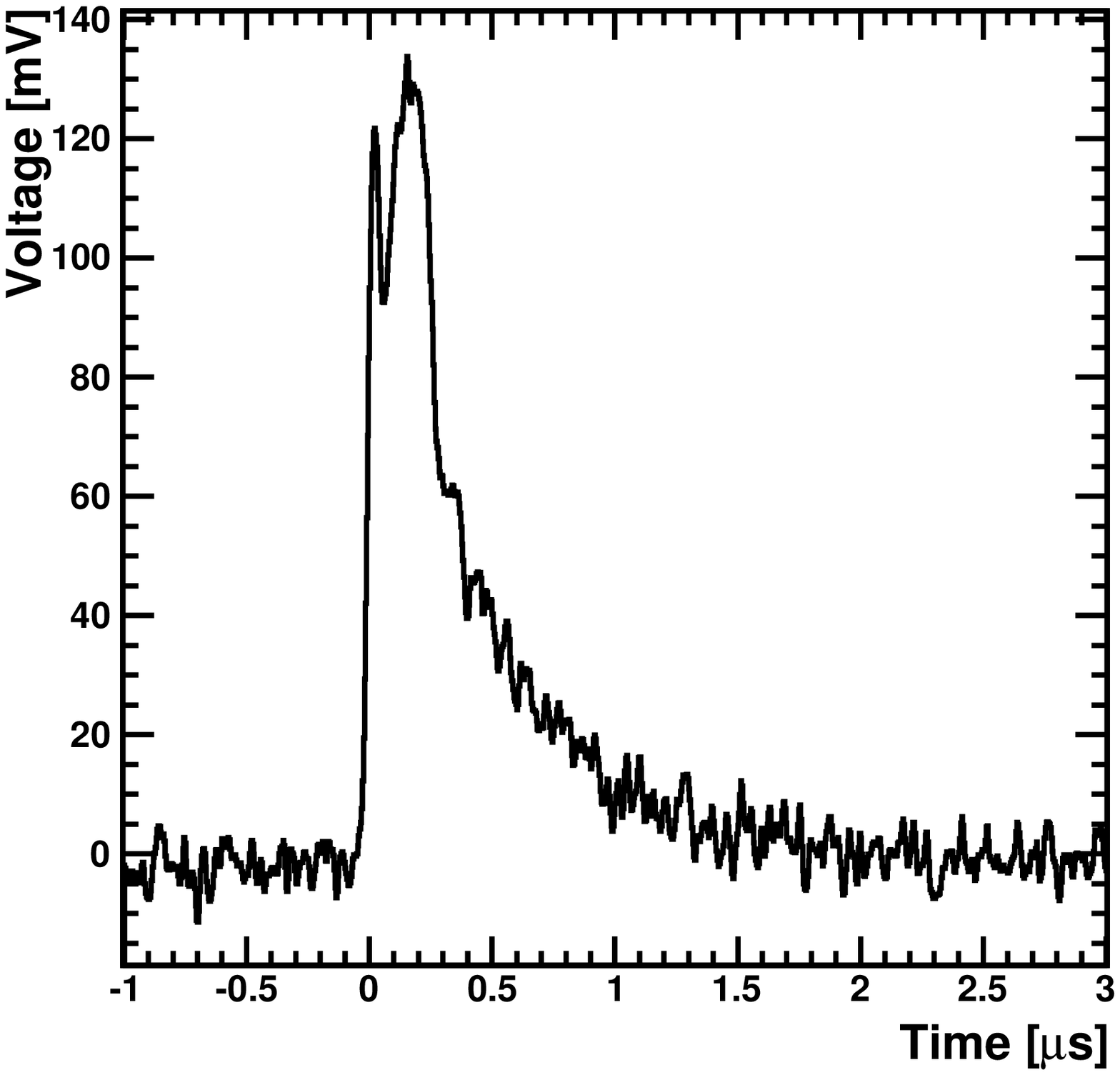}
    \includegraphics[width=0.45\textwidth]{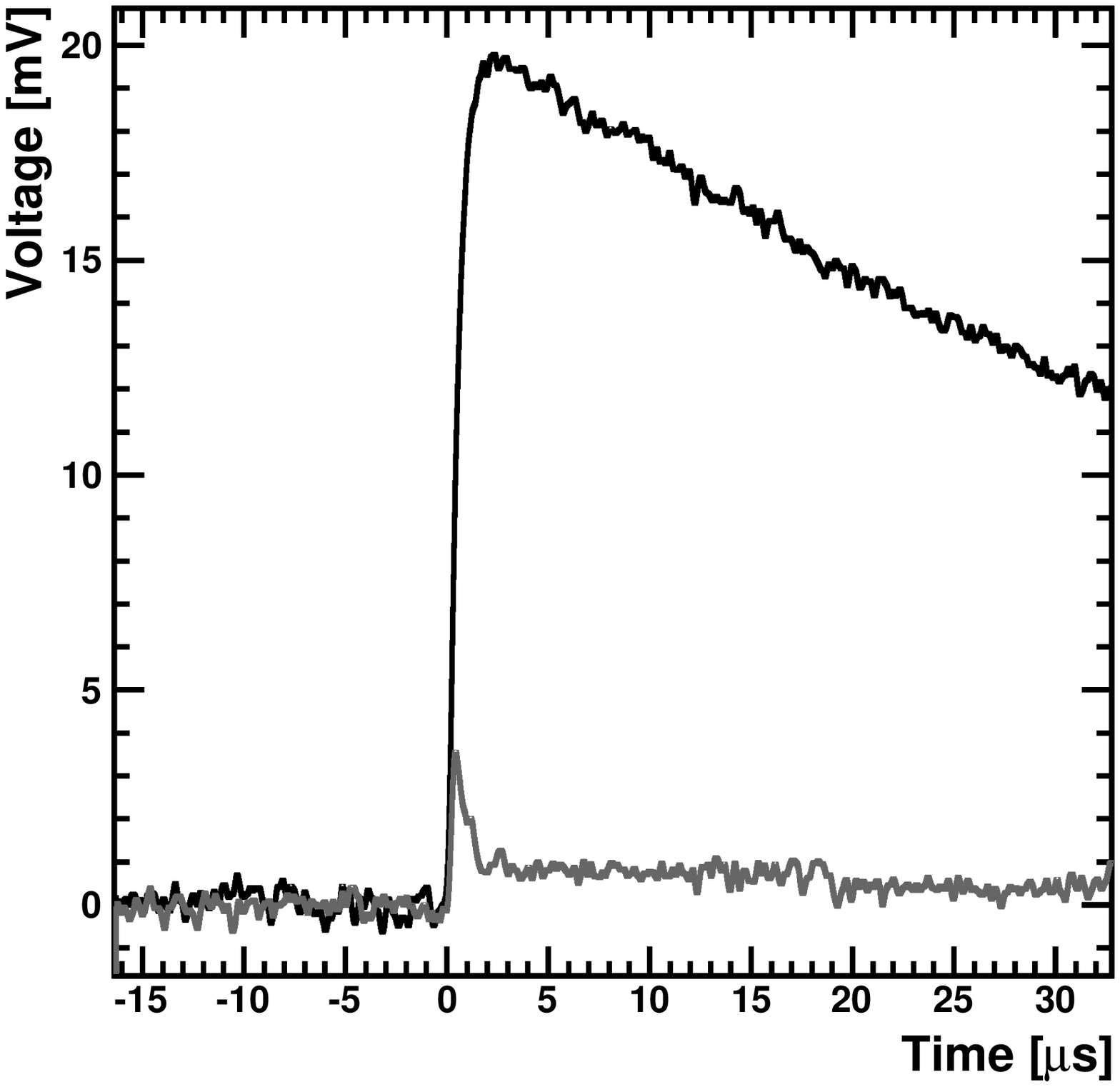}
   \label{fig:subfig1}  
   }

   \subfigure[60~keV$_\alpha$ $e^{-}$ recoil, with signal from veto electrode]{
     \includegraphics[width=0.45\textwidth]{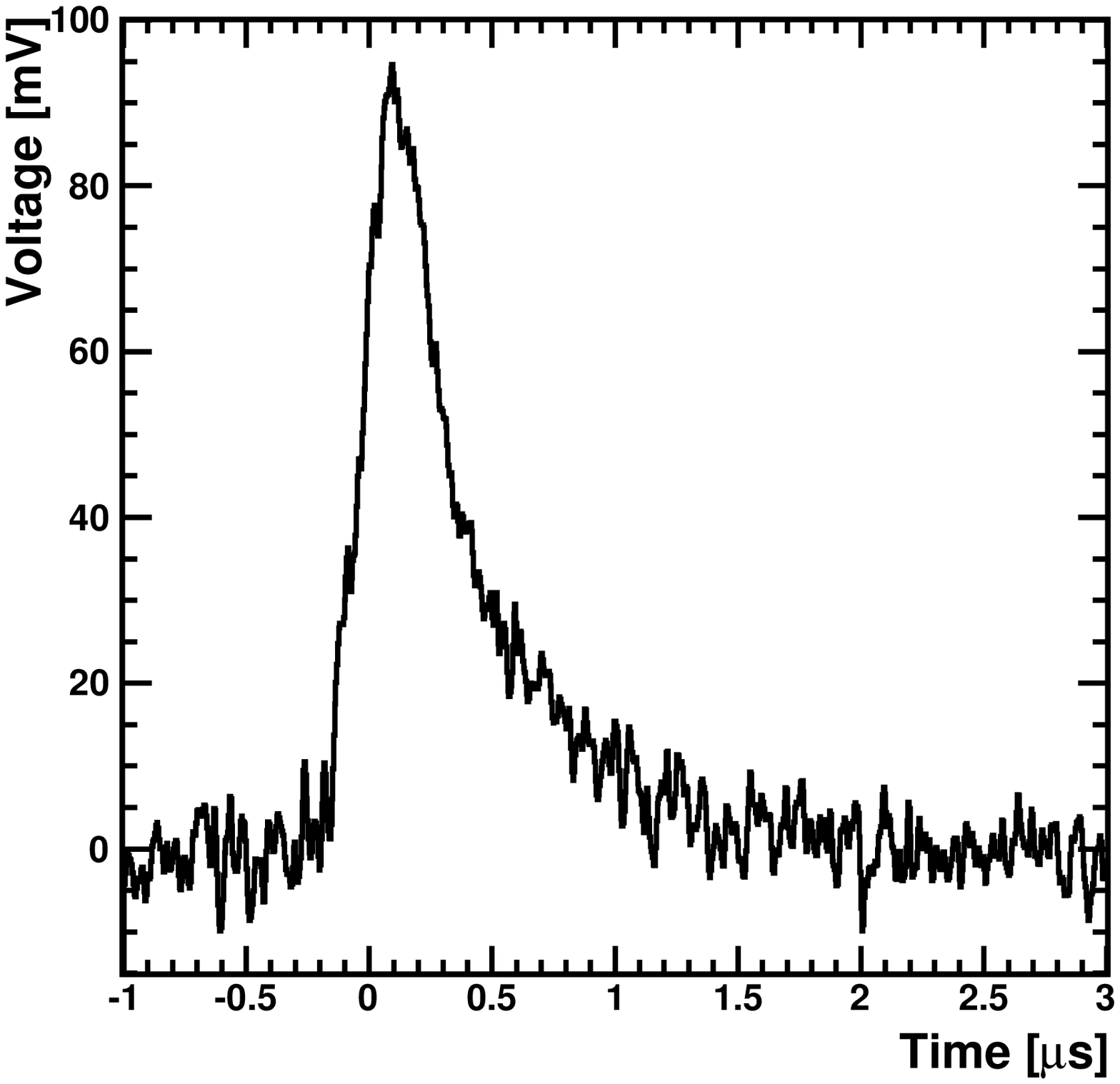}
     \includegraphics[width=0.45\textwidth]{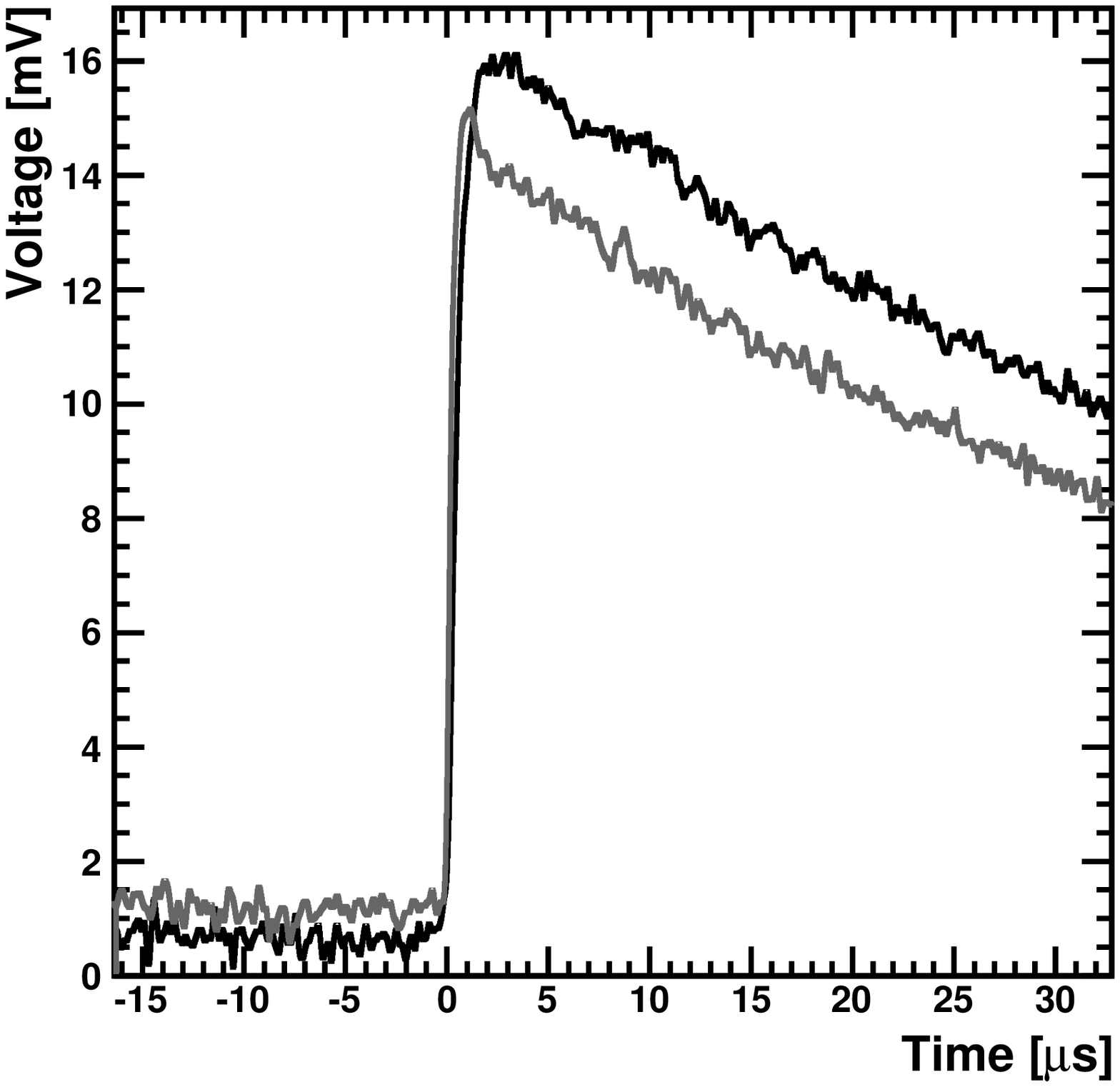}
     \label{fig:subfig2}  
   }

   \caption{Example smoothed traces of (a) an $\alpha$ track and (b) a vetoed electronic recoil.  Note the two-peaked structure of the mesh pulse in (a), which is absent in (b).}
   \label{fig:examples}
\end{figure}

The amplification mesh is read out through a Route2Electronics HS-AMP-CF preamplifier \cite{R2E}. 
This preamplifier has a gain of 80 and a rise time of roughly 1~ns when not connected to the detector.  A 30 $\Omega$ resistor connects the mesh and preamplifier input to ground so that the output is proportional to the induced current from charged particles drifting in the amplification region. 
Most ionization in the amplification region happens very near the anode so the electrons drift quickly over a very short distance while the ions drift more slowly over a longer distance, from the anode toward the amplification mesh.
The induced current of the electrons from a single electron is a peak that decays within several nanoseconds while the ions create a broader but smaller shoulder.
Low-energy nuclear recoils create compact ionization trails, with a range along the drift direction ($\Delta z$) of no more than a few millimeters. Because of the electron drift velocity of approximately 10~cm/$\mu$s \cite{Christophorou}, all primary ionization electrons reach the amplification region within a period of several tens of nanoseconds. The electron signals of the avalanches add together to create a fast rising edge and a sharp peak in the current signal.
The ions drifting in the amplification region then create a second, broader peak. This shape can be seen in Fig.~\ref{fig:subfig1}.  
In tracks with large $\Delta z$, such as most electronic recoils and minimum ionizing particles, the spread in time of primary ionization electrons entering the amplification region is long compared to the detector response of an electron avalanche, and the resulting pulse is characterized by a single broad peak from both the electrons and ions in the avalanche.  
Pulse shape analysis in the charge channels (see Fig. \ref{fig:examples}) provides powerful discrimination between nuclear and electronic recoils and can strongly suppress such backgrounds in a rare-event search.

All charge signals are digitized by AlazarTech ATS860 8-bit PCI digitizers using a sampling rate of 250~MHz.  The digitizer bandwidth extends up to 65 MHz. A total of 12288 samples (49.2~$\mu$s) are saved with each trace, including 4096 pre-trigger samples (16.4~$\mu$s).
Charge events are triggered on the rising edge of either the mesh channel signal at 75~mV or the central anode signal at 10~mV. This is sufficient to obtain a high expected efficiency for $E>$30 keV$_\alpha$.  
During event readout the CCD is exposed for 1~s, while the digitizers collect charge triggers.  After each exposure the image and charge triggers (if any) are written to file for later processing and analysis.  
A camera shutter is not used so tracks occurring during the shifting and digitization of CCD pixels, a process taking approximately 200~ms, are seen in the CCD data but not in the charge data.

\section{Event Reconstruction}
\label{sec:reconstruction}

\subsection{CCD Tracks}
Recoil candidates are selected using the energy, range and several geometric moments computed from the CCD image.
In order to maximize the recoil analysis efficiency, less stringent cuts were used than those described in \cite{10LPaper}.  The CCD selection cuts are described in detail in Section \ref{sec:ccdrej}. In Monte Carlo studies, the CCD track finding and cuts achieve over $70\%$ efficiency for ionization yields greater than 40~keV$_\alpha$ and over 90$\%$ for ionization yields greater than 50~keV$_\alpha$.

\subsection{Anode and Veto Signals}

The anode and veto signals are first smoothed using a Gaussian convolution with $\sigma=80$~ns.  The typical rise time of a pulse in these channels is 1~$\mu$s, so the smoothing reduces the noise while having little effect on the pulse shape.  The pre-trigger region is then used to determine the baseline voltage and noise RMS.  The pulse shape is characterized by its peak voltage and time and the times on both the rising and falling edge where the pulse reaches 10\%, 25\%, 50\%, 75\% and 90\% of the baseline-subtracted peak height.

\subsection{Mesh Signals}

For the mesh channel, a Gaussian convolution with $\sigma=6$~ns is used to reduce the noise. After smoothing, the 25\% to 75\% rise time of the initial sharp edge of the electron signal in the data used in this paper is always greater than 15~ns. The broadening due to the smoothing here is no more than approximately 10\%. Other features are less sensitive to the smoothing algorithm. 

\begin{figure}
\centering
\includegraphics[width=0.75\textwidth]{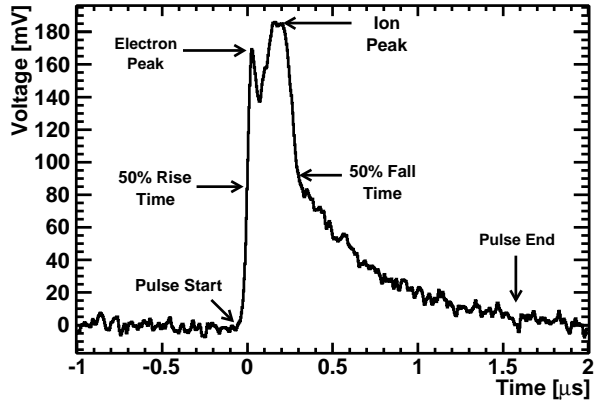}
\caption{Annotated mesh pulse of a 115~keV$_\alpha$ $\alpha$ particle. A Gaussian convolution has been used to reduce the noise prior to pulse shape characterization. The pulse start and end points are the nearest baseline crossings to the peak. Rise times are calculated from the electron peak and fall times are calculated from the ion peak. For anode and veto pulses, only a single peak is identified.}
\label{fig:annotated}
\end{figure}

As with the anode and veto channels, the baseline voltage and noise RMS are calculated from the pre-trigger samples. The initial electron peak of a nuclear recoil candidate is identified as the first peak in a pulse with a height greater than 50\% of the total peak height.  The ion peak is defined as the largest peak in the pulse separated from the electron peak by more than 50~ns. The lowest point between the two peaks is also measured.  The pulse shape is also characterized by similar rise and fall time variables as for the other channels. Rise times are calculated between the initial pulse baseline crossing and the electron peak and fall times between the ion peak and the baseline crossing at the end of the pulse. If only a single peak is found, rise and fall times are calculated from that peak. Single-peaked events occur rarely because the reconstruction algorithm will typically identify a small noise fluctuation as an additional peak if no clear second peak is present. Fig. \ref{fig:annotated} shows an annotated example of a mesh pulse from an $\alpha$ particle.

\subsection{Nuclear Recoil Selection Criteria}
\label{sec:select}
Selection criteria for nuclear recoils were determined by placing a $^{241}$Am $\alpha$ source above the cathode mesh and outside the active volume of the detector.  Most of the $\alpha$ energy is lost before crossing through the cathode into the active volume, so that $\alpha$ tracks of tens to a few hundred keV in energy are measured. 

Three sets of cuts are used to suppress background charge traces: 
(1) removal of triggers on noise and other pathological events;
(2) removal of signals from tracks passing over the veto region; and
(3) removal of events identified as electronic recoils (Table \ref{table:cuts}). 

The electronic noise-reduction cuts remove events with anomalous baseline voltages or noise RMS voltages. Any charge triggers due to noise on one of the channels are removed using cuts on the pulse rise and fall times. The analysis also checks that the full mesh pulse is included in the saved waveform and that the mesh and anode pulses are correlated in time. Charge events that saturate the digitizer for one of the charge channels are removed.

\begin{table}
\begin{tabular}{l l}
\hline\hline
Variable & Description \\
\hline
$V_{A}$ & Peak anode channel voltage \\
$V_{V}$ & Peak veto channel voltage\\
$T_{R}^{V}$ & 25\% to 90\% rise time of the veto channel\\
$T_{R}^{M}$ & 25\% to 75\% rise time of the mesh channel\\
$V_{e}/V_{A}$ & Ratio of the mesh electron peak and the anode peak\\
$V_i/V_A$ & Ratio of the mesh ion peak and the anode channel peak\\
\hline\hline
\end{tabular}
\caption{Description of reconstructed pulse shape variables used in analysis cuts.}
\label{table:cuts}
\end{table}

Events in the central anode region induce small pulses in the veto channel as well. Events actually passing through the veto region are rejected by requiring that $V_A > 4V_V$ and that $T_R^V<400$~ns. Pulses with longer rise times show the characteristic shape of electron avalanches occurring in the amplification region above the veto channel and are rejected with the latter cut.
Finally, a small population of pulses showing pileup effects are rejected by requiring time coincidence between the veto pulse and the mesh pulse. This population is very rare and typically only occurs when a radioactive source is placed inside the detector.

\begin{figure}
\centering
\includegraphics[width=0.75\textwidth]{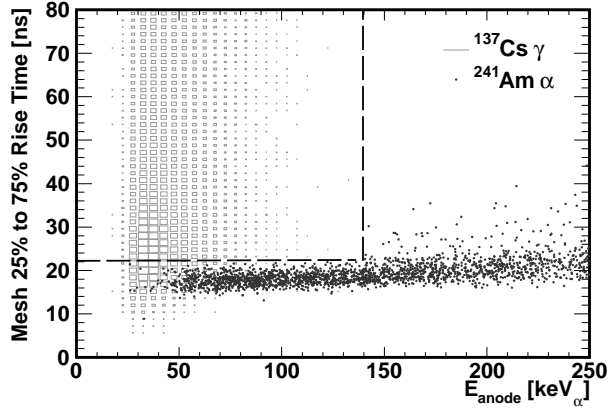}
\caption{Mesh signal rise time of both electronic recoils and $\alpha$ tracks. The dashed line represents the selection cut used in this analysis to remove electrons.}
\label{fig:risetime}
\end{figure}

\begin{figure}
\centering
\includegraphics[width=0.75\textwidth]{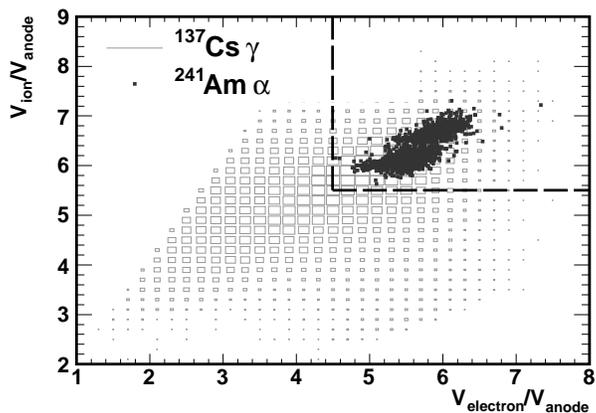}
\caption{Ratio of mesh signal peaks to anode signal peak of electronic recoils and $\alpha$ tracks. The dashed line represents the selection cuts used to remove electrons.}
\label{fig:peakratios}
\end{figure}

Nuclear recoil candidates are then identified using the shape of the mesh pulse.  
The rise time is due to a combination of electron longitudinal diffusion during drift, electronics response and recoil $\Delta z$.  
Nuclear recoil candidates have $T_R^M< 22$~ns for ionization yields of less than 125~keV$_\alpha$ (Fig. \ref{fig:risetime}).  
The compact tracks from nuclear recoils lead to generally larger but narrower peaks compared to electronic recoils of the same energy (Fig. \ref{fig:peakratios}).  
In this analysis, nuclear recoil candidates have $V_e/V_A>4.5$ and $V_i/V_A>5.5$. 

\begin{figure}
\centering
\includegraphics[width=0.75\textwidth]{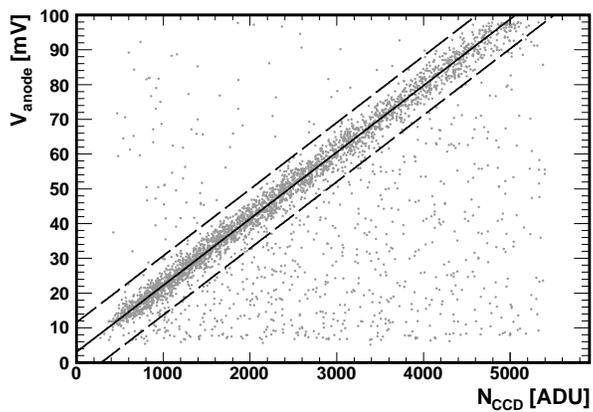}
\caption{Energy of $^{241}$Am track as measured by charge readout on the anode $V_{\mathrm{anode}}$ and by the CCD $N_{\mathrm{CCD}}$.  Solid line: best fit. Dashed lines: Energy matching cut limits, 3.5$\sigma$ from best fit.}
\label{fig:chargeLightFit}
\end{figure}

Once a set of candidate nuclear recoil light signals (using the cuts described in Section \ref{sec:ccdrej}) and a set of candidate nuclear recoil charge signals are identified, the event reconstruction attempts to match each track to its corresponding charge signal.  
To do this, all possible charge-light signal pairs are considered. 
The best match according to the relative charge-light energy calibration of $V_{\mathrm{anode}}[mV] = (3.07\pm0.04)+(0.01916\pm0.00002) N_{\mathrm{CCD}}[ADU]$ (Fig. \ref{fig:chargeLightFit}), where $V_{\mathrm{anode}}$ is the peak height of the anode signal and $N_{\mathrm{CCD}}$ is the total number of ADU in the CCD track, is chosen.  The match is accepted if the anode signal is less than 8.5 mV (3.5$\sigma$) from the value estimated from the measured light signal.

At the energies measured here, the trail of ionization left by $\alpha$ particles is much longer than that left by fluorine and carbon recoils of the same energy. The $\alpha$ particles in the calibration data enter the field cage at a mean polar angle of approximately 45$^{\circ}$ and drift across the full length of the drift cage (see the source placement in Fig. \ref{Detector_Drawing}), so the measured parts of the $\alpha$ tracks have more diffusion and $\Delta z$ comparable to or longer than nuclear recoils oriented exactly along the drift ($z$)direction.  Because of this, the cuts set using the $^{241}$Am data will be valid for carbon and fluorine recoils as well. The efficiency of the charge cuts and charge-light energy matching for tracks identified in the CCD analysis was measured using the $\alpha$ data to be greater than 90\% for $40~\rm{keV}_\alpha<E<200~\rm{keV}_{\alpha}$.

\begin{figure}
\centering
\includegraphics[width=0.75\textwidth]{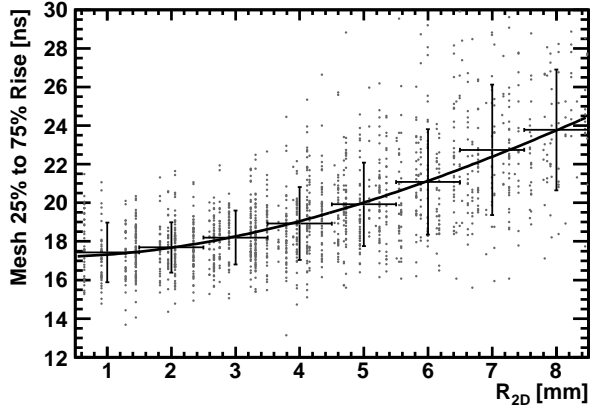}
\caption{25\%-75\% rise time vs. $R_{2\mathrm{D}}$ of low energy $\alpha$ tracks.  The tracks are collimated with a mean angle of roughly 45$^\circ$, so $\Delta z$ increases with $R_{2\mathrm{D}}$. The fit is to $T(R_{2\mathrm{D}}) =\sqrt{ T_0^2 + m^2R_{2\mathrm{D}}^2}$, where $T_0=17.19\pm0.06$~ns is the contribution to the rise time from diffusion and $m=2.04\pm0.03$~ns/mm is the slope of the line at high $R_{2\mathrm{D}}$. 
The error bars here represent the 1$\sigma$ spread from the mean value.  
A large spread is expected due to both straggling effects and the initial energy and angle distribution of the source.}
\label{fig:deltaZ}
\end{figure}

The $\alpha$ data was also used to evaluate the capability of using the rise time to measure the recoil $\Delta z$.  
Because the mean polar angle of the $\alpha$ particles is roughly 45$^\circ$, the two-dimensional range ($R_{2\mathrm{D}}$) measured by the CCD is roughly proportional to $\Delta z$.  Due to straggling and imperfect collimation of the source, a range of $\alpha$ energies and incident angles is seen, so a wide range of $\Delta z$ values is expected for a given $R_{2\mathrm{D}}$. However, the mean value can still be used to determine a calibration between the rise time and $R_{2\mathrm{D}}$. 
Fig. \ref{fig:deltaZ} shows the result using the 25\% to 75\% rise time of the mesh signal. The proportionality constant between this rise time variable and $R_{2\mathrm{D}}$ is measured to be 2.04$\pm$0.03~ns/mm.

\section{Background Rejection}

\subsection{Electronic recoils}
\label{sec:emrej}
To determine the ability of a DMTPC detector to reject electronic recoils, a collimated 5~$\mu$Ci $^{137}$Cs source was placed above the cathode mesh, outside the fiducial volume of the detector.  The monoenergetic 660~keV $\gamma$ rays create a broad spectrum of electronic recoils across the entire energy range used for WIMP searches, while higher energy 512~keV and 1.2~MeV electrons from $\beta$ decays may also be detected.  

An average of 27 charge signals per 1~s exposure passed the noise and pathological event cuts during the $^{137}$Cs data run. 
Pile-up in the charge signals is negligible at this event rate. 
However, the signals of all recoils are accumulated into a CCD image, so that a region where several recoils 
overlap may be misidentified as a nuclear recoil, even if any single recoil would not be observable above the CCD read noise.

\begin{figure}
\centering
\includegraphics[width=0.75\textwidth]{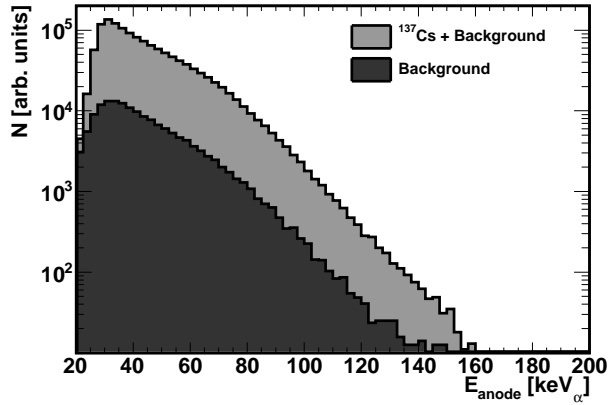}
\caption{Light: Energy spectrum measured by the central anode channel in the $^{137}$Cs run.  Dark: Measured background spectrum scaled to match the live time of the $^{137}$Cs run. The charge trigger efficiency decreases below 30~keV$_\alpha$ leading to the apparent peak in the spectrum.}
\label{fig:gammaspec}
\end{figure}

A separate run with no sources inside the chamber was taken to measure the expected background spectrum. 
After removing events less than 3 seconds following a spark on the anode, a total of 51053 and 32568 one-second exposures were taken in the $^{137}$Cs and background runs, respectively.  
The distribution of all charge events passing the initial set of noise and pathological-event cuts is used to determine the number of electronic recoils, as the trigger rate is dominated by electrons and minimum ionizing particles.
The energy spectra of the two runs, after applying only the quality cuts identifying valid charge events, are shown in Figure \ref{fig:gammaspec}.  

In Monte Carlo studies, the CCD analysis used in this study has an efficiency of over 90\% for nuclear recoils with energies between 40~keV$_\alpha$ and 200~keV$_\alpha$.  
In this energy range, a total of 679919 charge events were measured in the $^{137}$Cs run with an expected background of $77310\pm350$. 
 In the CCD analysis 20 tracks in the background run and 63 in the $^{137}$Cs run pass all cuts without considering any charge cuts.  
 The higher event rate in the $^{137}$Cs run provides evidence for possible light pileup events; many CCD tracks occur near the track-finding energy threshold.  The number of charge traces passing each set of cuts is given in Table~\ref{table:rates}.

\begin{table}
\centering
\begin{tabular}{llll}
\hline \hline
 & $^{137}$Cs& Background & Bkg.-Subtracted $^{137}$Cs \\
\hline
Noise Cuts & 679939 & 49339&602630$\pm$450\\
+Veto Cuts & 38499&3891&32400$\pm$130\\
+$e^-$ Cuts& 255& 35&199$\pm$12\\
+Light Matching &5 & 3&-1.27$^{+3.7}_{-3.3}$\\
\hline \hline
\end{tabular}
\caption{Number of charge triggers with 40~keV$_\alpha$ $<E_{anode}<$ 200~keV$_\alpha$ that pass the specified cuts. The background subtraction includes the uncertainty of the background measurement.}
\label{table:rates}
\end{table}

The camera only images part of the amplification plane while the charge channels read out the entire plane. A background-subtracted sum of images taken during the $^{137}$Cs run is used to estimate the fraction of electronic recoil events occurring within the region viewed by the CCD. 
A fit of this sum to a two-dimensional Gaussian distribution shows that roughly 68\% of all ionization from electronic recoils occurs in this region. 
This value is known to within roughly 15\%. It is expected that approximately the same percentage of recoils also occur in the region read out by the CCD. Using this number and subtracting the expected background, 409770$\pm$430 electronic recoils occurred in the fiducial volume out of a total of 602600$\pm$350 with $40~\rm{keV}_\alpha<E<200~\rm{keV}_\alpha$.  
Using these numbers, the CCD analysis alone achieves an electron rejection factor of $7.3\times10^{-5}$ in this energy range. The charge analysis includes all recoils occurring in the field cage and rejects electron recoils at the level of $3.3\times10^{-4}$.

\begin{figure}
\centering
\includegraphics[width=0.75\textwidth]{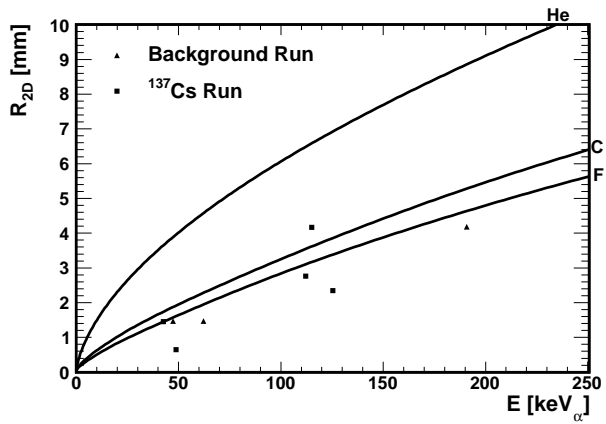}
\caption{2-dimensional range vs energy of events in the $^{137}$Cs and background runs that pass all cuts.  Solid lines are SRIM predictions for carbon, fluorine and helium tracks. The measured range is two-dimensional, while the SRIM curves give three-dimensional ranges so the measured points will generally fall below the SRIM prediction.}
\label{fig:rangeEnergy}
\end{figure}

A combined analysis using both charge and CCD cuts as well as charge-light energy matching yields 5 events in the 
$^{137}$Cs run and 3 events in the background run. Including the difference in total exposure time and the large 
uncertainties on both the background and combined background and signal distributions, a 90\% confidence level 
upper limit on the electron recoil rejection factor of $1.1\times10^{-5}$ is reached in the 
40~keV$_\alpha$ to 200~keV$_\alpha$ energy range of the recoil spectrum of $^{137}$Cs $\gamma$ rays. From the detection efficiencies determined by Monte Carlo studies and the $\alpha$ source measurements, the detection efficiency for nuclear recoils using a combined CCD and charge analysis is determined to be over 63\% for energies greater than 40 keV$_\alpha$ and over 80\% for energies greater than 50 keV$_\alpha$.   
A plot of the two-dimensional range as measured by the CCD against the recoil energy as measured by the anode channel shows
(Fig. \ref{fig:rangeEnergy}) that the passing events appear near the predicted three-dimensional range for nuclear recoils from SRIM. 
The two-dimensional range is shorter than the full three-dimensional range, so nuclear recoils are expected in a broad band with shorter ranges than the SRIM prediction.  
Furthermore, the peak pixel values for the tracks are well above the threshold used in track finding,
indicating that the events are likely to be nuclear recoil or $\alpha$ backgrounds rather than signal pileup from electrons.
If these events were excluded as likely nuclear recoils,
  the result would be statistics-limited at $5.6\times10^{-6}$.

\subsection{CCD Backgrounds}
\label{sec:ccdrej}

Several classes of CCD background events appear in our detectors and have previously been eliminated through the use of CCD track selection cuts.  
While these cuts are generally able to identify and remove most such events, the charge readout analysis is also useful at eliminating these events. 
With the addition of the charge readout analysis, much less restrictive cuts can be used in the CCD analysis to maximize the efficiency for nuclear recoils.  
These background events include (1) residual bulk images, (2) intermittent hot pixels and (3) noise events.  CCD backgrounds do not originate from ionization in the gas volume and so should have no associated charge signal.  

Residual bulk images (RBI's) generally occur in our detectors when a spark inside the amplification region causes a great deal of light to hit the CCD.  A fraction of the longer wavelength light penetrates deeply enough into the silicon to generate photoelectrons in the depletion region.  
With little to no electric field in this region, the free charge carriers must diffuse via thermal motion into the pixel potential wells. This can  take several minutes \cite{CCDbook}.  
The strength of the RBI signal is proportional to the exposure time and can sometimes mimic low energy nuclear recoils. 
Residual bulk images occur in front-illuminated CCDs such as the Kodak KAF-1001 chip used in the Apogee Alta U6 camera but not in back-illuminated CCDs. 

Hot pixels and ionization from recoils occurring directly in the CCD chip are another source of background events found in the CCD event reconstruction.  
Hot pixels are a readout artifact where a single pixel is read out as containing an anomalously large amount of charge. 
Particles such as muons and $\gamma$ rays incident on the CCD can also leave ionization inside the CCD silicon that will appear in one or more pixels.  
Such tracks typically leave much more charge per pixel than light from avalanches inside the detector gas volume.   

While the cuts described in \cite{10LPaper} are able to remove most of these CCD artifacts, they also can significantly reduce the reconstruction efficiency for low energy nuclear recoils, where nuclear recoils are more difficult to distinguish from the background noise of the CCD.  
Requiring coincidence between the light and charge signals provides additional rejection power of these types of events.  The CCD analysis is less sensitive to recoils than the charge analysis due to the favorable signal-to-noise ratio of the charge channels compared to the CCD.

In addition to these CCD artifacts another CCD-related background is ``out-of-time" events, which are recoils and $\alpha$ decays occurring during event readout.  
When the CCD is being read out, charge is shifted from one pixel to another as the charge from each pixel is digitized.  
A camera shutter is not used during data acquisition so any scintillation occurring during the 0.2-0.3 s readout time will appear shifted in the digitized image from its true position.  
The scintillation light from an $\alpha$ particle depositing only a small fraction of its energy in the imaged region can then be shifted toward the middle of the image, where it might resemble a nuclear recoil. 
These events are very difficult to identify with the CCD analysis.  
The charge channels do not collect data during this readout time, so a coincident charge signal will not appear for these shifted events.

\begin{table}
\begin{tabular}{llll}
\hline \hline
Event type        &     Before &  After      & Reduction [\%]\\
from CCD analysis & charge cuts& charge cuts \\
\hline
RBI &1332&7 & $99.5\pm0.2$\\
Hot pixel/CCD Si Recoil&1246 &11 &$99.1\pm0.3$ \\
Edge Crossing ($\alpha$) &17 &0 & $100_{-9}^{+0}$\\
Nuclear recoil/Out-of-time $\alpha$ &20 &5 &$75\pm11$\\
\hline
All Tracks &2615 & 23 &$99.1\pm0.2$\\
\hline \hline
\end{tabular}
\caption{Number of CCD events of different types found before and after applying charge cuts, 25 keV$_\alpha$ $< E_{CCD} < 400$ keV$_\alpha$.  The RBI and hot pixel/CCD Si recoil events are false coincidences between low energy charge signals and CCD artifacts with low apparent light signals.  }
\label{table:ccdrej}
\end{table}

To evaluate the ability of the charge readout in removing these CCD background events in the energy range 25 keV$_\alpha<E_{\mathrm{CCD}}<400$ keV$_\alpha$, a separate analysis was performed on the data from the background run described in Sec. \ref{sec:emrej}. 
A reduced set of CCD cuts identical to those used in Sec. \ref{sec:emrej} is used to define different classes of CCD artifacts and to determine the fraction removed by requiring charge-light coincidence. 
RBIs are defined as having at least two CCD tracks occurring within 10 pixels (1.6 mm) of one another within a single 1000 event run.  
Hot pixels, ionization events in the CCD chip and noise events are identified by three cuts. They 1) have a maximum pixel value greater than 500 ADU (38 keV), 2) include a pixel containing more than 30\% of the total light of the reconstructed track, or 3) do not include enough pixels far enough above background to reconstruct a nonzero range.
Tracks passing these cuts but touching the edge of the image are typically $\alpha$ decays and are removed as well.  
Any remaining tracks are tagged as nuclear recoil candidates or out-of-time partial $\alpha$ events. 

Requiring all CCD events to have a coincident charge signal removes more than 99\% of the RBI, noise, CCD Si track, and hot pixel events before applying any of the CCD cuts.  
The few events of these types that did pass were due to false coincidences between CCD and charge signals with very low energies.  
All edge crossing events were removed, as would be expected since the CCD only measures part of these events.
Finally, $75\pm11\%$ of the events tagged as possible nuclear recoils in the CCD-only analysis were also rejected, leaving 5 nuclear recoil candidates having both charge and light signals.  This indicates that many of the events identified in the CCD analysis are likely either out-of-time nuclear recoils and $\alpha$ decays or events from the other classes that were not identified by the selection cuts (Table. \ref{table:ccdrej}).

\section{Discussion and Conclusions}

This work has demonstrated the ability to reject electronic recoils by a factor of  $1.1 \times 10^{-5}$ at 90\% C.L. level for electron recoils generated from a $^{137}$Cs $\gamma$-ray source with energies between $40$~keV$_\alpha$ and 200~keV$_\alpha$ (or between 80~keV$_r$ and 300~keV$_r$ for fluorine recoils).  Neither the charge nor the CCD analysis is completely efficient at removing electrons when considered independently in this analysis, but the combined result is significantly stronger. In an underground WIMP search, the CCD analysis is expected to be much more effective due to the greatly reduced recoil multiplicity per exposure compared to the $^{137}$Cs run shown here. Stronger selection cuts will also further enhance the ability of the CCD analysis to reject electrons. The charge-light matching will also be much more effective as the chance of finding a false coincidence will be greatly reduced with the lower event rate in a source-free run.

It is also expected that the CCD and charge analyses are most sensitive to different event topologies. The charge analysis is most effective at rejecting electronic recoils with large $\Delta z$, while the CCD analysis is most efficient at rejecting electrons with large projected two-dimensional range, perpendicular to $\Delta z$. With a rejection power in this simple analysis of $7.3\times10^{-5}$ in the CCD, $3.3\times10^{-4}$ in charge, and additional rejection power gained by requiring charge-light energy matching a rejection factor of order $10^{-8}$ could be achieved if the CCD and charge analyses are in fact relatively uncorrelated. A much longer run would be required to attempt to confirm such a hypothesis. Even with this rejection power, the electron rejection analysis can be improved. Stronger cuts on recoils found in the CCD have already been used by the DMTPC collaboration in WIMP searches, although these also reduce the detector efficiency for signal events. Further refinements in the charge reconstruction and selection cuts are also expected in future analyses.

The addition of the charge signal analysis strengthens the ability of DMTPC detectors to identify and remove CCD artifacts from the analysis by removing more than 99\% of such events before applying CCD-based nuclear recoil selection cuts.  The charge analysis also allows for the identification and removal of events occurring during CCD readout, which can mimic the signal of low-energy nuclear recoils but cannot be easily identified with the CCD-only analyses used previously by the DMTPC collaboration.

The ability to reject electronic recoils with high efficiency and without significantly reducing the detection efficiency of nuclear recoils suggests that electronic recoils are not expected to be a significant source of background events for the target energy range of DMTPC detectors for WIMP searches.  
\section*{Acknowledgments}
The DMTPC collaboration would like to acknowledge support by the U.S. Department of Energy (grant number
DE-FG02-05ER41360), the Advanced Detector Research Program of
the U.S. Department of Energy (contract number 6916448), as well as the
Reed Award Program, the Ferry Fund, the Pappalardo Fellowship program,
the MIT Kavli Institute for Astrophysics and Space Research, the MIT
Bates Research and Engineering Center, and the Physics Department at the
Massachusetts Institute of Technology. We would like to thank Mike
Grossman for valuable technical assistance.
\bibliographystyle{unsrt}
\bibliography{gammaPaper}

\end{document}